\documentclass[12pt]{article}
\usepackage{lineno,hyperref}

\usepackage{scicite,times}
\newenvironment{sciabstract}{%
\begin{quote} \bf}
{\end{quote}}

\usepackage{soul}
\usepackage{color,xcolor}
\usepackage{amsmath,mathtools,amsfonts,amssymb,mathptmx}
\usepackage{graphicx}
\usepackage{txfonts}
\usepackage{subcaption,caption}
\usepackage{multirow}
\usepackage{tabularx}
\usepackage[export]{adjustbox}

\hypersetup{
    colorlinks,
    linkcolor={red!90!red},
    citecolor={blue!50!black},
    urlcolor={cyan!100!black}
}

\usepackage{url}

\graphicspath{{Final_Figures/}}

\title{Higher order organizational features can distinguish protein interaction 
networks of disease classes: a case study of neoplasms and neurological 
diseases}

\author
{Vikram Singh,$^{1\dagger}$ Vikram Singh,$^{1\ast}$\\
\\
\normalsize{$^{1}$Centre for Computational Biology and Bioinformatics, Central 
University of Himahcal Pradesh,}\\
\normalsize{Dharamshala,  Himahcal Pradesh, 176206, India} \\
\\
\normalsize{$^\ast$To whom correspondence should be addressed; E-mail: 
vikramsingh@cuhimachal.ac.in}
}

\date{}

\begin{document} 
\baselineskip24pt
\maketitle
\begin{sciabstract}
Neoplasms (NPs) and neurological diseases and disorders (NDDs) are 
amongst the major classes of diseases underlying deaths of a 
disproportionate number of people worldwide. To determine if there exist some 
distinctive features in the local wiring patterns of protein interactions 
emerging at the onset of a disease belonging to either of these two classes, we 
examined 112 and 175 protein interaction networks belonging to NPs and NDDs, 
respectively. 
Orbit usage profiles (OUPs) for each of these networks were enumerated by 
investigating the networks' local topology. 
56 non-redundant OUPs (nrOUPs) were derived and  
used as network features for classification between these two disease classes. 
Four machine learning classifiers, namely, k-nearest neighbour (KNN), support
vector machine (SVM), deep neural network (DNN), random forest (RF) were 
trained on these data. DNN obtained the greatest average AUPRC (0.988) among 
these classifiers. 
DNNs developed on node2vec and the proposed nrOUPs embeddings were compared 
using 5-fold cross validation on the basis of average values of the six  
of performance measures, viz., AUPRC, Accuracy, Sensitivity, Specificity, 
Precision and MCC. 
It was found that nrOUPs based classifier performed better in all of these 
six performance measures. 

\end{sciabstract}

\section{Introduction}
Understanding complex human diseases require linking disease phenotypes to their 
underlying molecular mechanisms and genetic components 
\cite{weatherall2001phenotype}. With the advent of the biological big-data age, 
characterising the molecular or genetic elements of diseases has become equally 
crucial as characterising them based on their pathological and clinical 
manifestations \cite{oti2008phenome}. Next-generation sequencing (NGS) and 
genome-wide association studies (GWAS) have facilitated the collection of many 
disease-gene associations \cite{tam2019benefits}. Parallelly, extensive protein 
interaction maps have been constructed due to recent advances in high-throughput 
proteomic technologies \cite{uniprot2021uniprot}. An organism's genes are 
involved in several vital functions, including regulating cellular processes. 
The complex disease phenotypes of living beings arise due to the disintegration 
of the functional connections between genes that are intricately strung 
together \cite{barabasi2011network}. Since then, researchers in the biological 
sciences have used a wide range of experimental and computational methods to 
define the relationships between genes and the proteins they produce. 
Extensive population studies are being conducted to identify the genes 
implicated in common, multifactorial diseases and link the molecular pathology 
of simple monogenic disorders to their related clinical manifestations 
\cite{elliott2018genome}. The explanation of monogenic Mendelian inheritance 
patterns may aid the identification of pathogenic processes in complicated 
diseases. It is common for numerous genes to be involved in a complex disease, 
but pinpointing the exact genes responsible has proven difficult. The 
importance of understanding phenotype-genotype correlations in devising 
approaches to gene therapy is highlighted by the need for increased clinical 
research involving vast numbers of patients.

Network medicine is a new discipline that uses network science methods to 
extract meaning from massive omics datasets on molecular diseases. The 
development of genotype-based disease networks was an early attempt to 
investigate the systemic significance of disease-gene associations from a 
network perspective \cite{barabasi2011network}. These networks are valuable for 
illuminating the global organisation of diseases around functional modules and 
for deducing comorbidity relations between diseases \cite{hu2016network}. 
Many subsequent studies have sought to expand upon the foundation laid by these 
early works by examining links between diseases 
\cite{goh2012exploring,xiang2022biomedical}. For instance, random walk 
algorithm has been exploited to discover previously unknown associations 
between diseases in the protein-protein interaction network 
\cite{suratanee2015dda}. Similarly, differential coexpression analysis has been 
used to determine the degree of similarity between diseases and to reveal their 
shared molecular mechanisms, resulting in the discovery of novel interactions 
between diseases with previously unknown molecular mechanisms 
\cite{yang2015human}.

It has recently been established that the molecular architecture of human 
diseasome is modular \cite{oti2007modular}. Several genetic diseases are 
traced back to changes in genes that are all part of the same cellular pathway, 
molecular complex, or functional module, lending credence to this line of 
thinking \cite{joenje2001emerging}. It has been demonstrated, for instance, 
that proteins encoded by genes linked to the same disease are more likely to 
interact with one another than with proteins from other diseases 
\cite{goh2007human}. Moreover, in recent years a large volume of research has 
been focused on quantifying the similarities and differences among different 
disease classes using various similarity measures ranging from text mining to 
network topology. Various methods have been used to explore the connections 
between diseases. The most trustworthy approaches to identifying causal 
relationships between genes and diseases are genome-wide association studies 
(GWAS). Using GWAS data, researchers in several studies uncovered previously 
unknown connections between seemingly unrelated disorders 
\cite{huang2009identifying}. Comparing disease ontology (DO) and gene ontology 
(GO) trees for semantic overlap is another method that has been exploited for 
the same \cite{mathur2012finding}. Moreover, these network-based methods have 
been used to study disease comorbidity and infer novel disease-associated 
genes. 

Cancer and nervous system diseases are two critical issues in public health and 
are responsible for a significant number of deaths all around the globe. Cancer 
is a complicated, polygenic disorder that results from mutations in many genes 
\cite{hanahan2022hallmarks}. Contrarily, some brain diseases have 
monogenic roots, and others are polygenic \cite{mccarroll2013progress}. Rapid 
advances in high throughput approaches for generating biological data have made 
it possible to use a systems-level, integrative approach to study complex 
diseases \cite{kar2009human,santiago2014network}. In this study, we attempt to 
characterise the higher-order organisational patterns underlying 220 protein 
interaction networks belonging to these two classes of diseases and use these 
patterns to classify different networks into their respective classes.

\section{Methods}

\subsection{Data acquisition and disease network construction}

DisGeNET maintains one the most extensive collection of genes associated with 
various monogenic, polygenic, rare diseases (related) and environmental traits 
\cite{pinero2020disgenet}. It combines data from various resources, including 
manually curated databases, text mining, GWAS datasets, and animal models, into 
one resource. The most authoritative disease-gene associations, \emph{i.e.} 
those with GDA scores greater than 0.1, for neoplasm (MeSH class C04),  nervous 
system diseases (C10), and two classes of psychiatry and psychology (F) MeSH 
superclass; including behavior and behavior mechanisms (F01), mental disorders 
(F03) were obtained from DisGeNET. Recent research from the Pan-Cancer 
Initiative \cite{weinstein2013cancer} have shown that tumours of various organs 
share molecular features, while tumours of the same tissue may have vastly 
distinct genetic features. Motivated with these results, we combined all the 
types of neoplasms in one group and combined C10, F01, and F03 in one group 
called the neurological diseases and disorders (NDDs). This grouping resulted 
in 331 diseases, 112 in the 
neoplasms (NPs) class and 219 in the NDDs class, associated with 10, 
880 
unique genes. These 10, 880 unique entrez gene ids belonging to the two disease 
classes were then mapped using UniPro Id mapper to 
identify their corresponding STRING ids. We could map 10, 624 gene ids, so we 
discarded the unmapped 256 gene ids. Moreover, the diseases with less than 100 
associated genes were also discarded. Out of 219 networks belonging to NDDs, 
 we were able to 
map at least 100 entrez gene ids to matching STRING ids for 175 networks. 
However, this was not the case for the neoplastic networks, each of which 
contains at least 100 matching STRING ids. Thus this brings down the total 
number of networks utilised throughout the study to 257. To reconstruct a PPI 
network, we used human protein-protein interaction (PPI) information from the 
STRING database version 11.5 \cite{szklarczyk2021string}. STRING includes both 
experimentally verified and computationally inferred PPIs, along with an 
associated confidence 
score for each physical and functional interaction. We constructed the human 
contact network by selecting only interactions with confidence scores of 700 or 
higher to ensure we got only trustworthy interactions. Then the largest 
connected PIN component was obtained, containing 252,833 connections among 
16,584 proteins.

\subsection{Differentially expressed orbits (DEOs) identification}

To construct random ensembles, we followed the same methodology explained in 
Singh et al. \cite{singh2022characterizing}. Two types of random networks, Erdos 
Renyi and density dependent scale free models, were used to construct random 
ensembles corresponding to each disease network. Then orbit degree vector 
matrices for every network were obtained using the orbit counting algorithm 
(ORCA), which were further normalised to get orbit usage profiles (OUPs). 
Similarly, OUPs for both random ensembles were also enumerated, which were then 
compared with OUPs of disease networks to identify differentially expressed 
orbits (for complete methodology, please refer to Singh et al. 
\cite{singh2022characterizing}).

\subsection{Preprocessing and classification of diseases}

Before being used for training classification models, all OUP vectors were 
subjected to removal of 17 redundant orbits \cite{yaverouglu2014revealing} and 
then each non-redundant OUP was appropriately normalised with min-max scaling. 
Formally, we used the following equation to normalise each N-dimensional feature 
vector \(x\):
\[\hat{x} = \frac{x - min(x)}{max(x) - min(x)}\],
ensuring \(0 \leq \hat{x} \leq 1 \forall i = 1, \ldots, N\), where \(N\) 
denotes OUP's dimensions. We trained the support vector machine (SVM), naïve 
Bayes (NB), K-nearest neighbour (KNN), random forest (RF), and multi-layered 
deep neural network (DNN) supervised machine learning algorithms on the 
normalized OUPs to classify them. Classification with SVM involves kernel 
tricks, which include mapping a lower-dimensional space to a higher-dimensional 
one. To enhance prediction performance, we opted for the radial basis function 
(RBF) kernel and tuned two additional parameters, kernel width (\(\gamma\)) and 
regularisation \(C\), between the ranges \(2^{-5}-2^{11}\) and 
\(2^{-13}-2^{3}\) respectively. The NB uses Bayes' theorem with the assumption 
that all possible feature pairs are uncorrelated with one another. Gaussian NB 
was used to construct the model for this investigation. In order to produce a 
prediction, KNN uses the geometric distance information of neighbours, which is 
a non-parametric technique. We choose five neighbours for this example. Random 
Forest (RF) is an ensemble method that constructs several decision trees by 
randomly sampling subspaces of feature vectors to make predictions. DNN 
implements one or more hidden layers (nonlinear approximators) between the input 
and output layers to make the classification. We have optimized the number of 
hidden layers, the number of units per layer, L2 regularization and solver 
parameters for DNN.

\subsection{Performance evaluation}

Five-fold cross-validation was used to compare the accuracy with which various 
classifiers predicted the training data. To ensure the predictors are reliable, 
randomly dividing the data into training and test sets, followed by model 
construction and evaluation, is performed a thousand times. Finally, the 
accuracy of the predictions was measured by four metrics: recall, specificity, 
accuracy, and Mathew correlation coefficient (MCC) which are defined as follows:

\[precision = \frac{TP}{TP + FP}\]
\[Recall = \frac{TP}{TP + FN}\]
\[Specificity = \frac{TN}{TN + FP}\]
\[MCC = \frac{TP \times TN - FP \times FN}{\sqrt{(TP + FP) \times (TP + FN) 
\times (TN + FP) \times (TN + FN)}}\],

where TP, TN, FP, and FN represent true positive, true negative, false positive 
and false negative protein pairs. Furthermore, this represents a binary 
classifier with unbalanced positive and negative data sets, so precision-recall 
curves were used, and the area under the precision-recall curve (AUPRC) was 
computed \cite{davis2006relationship}.

\section{Results and Discussion}
In this section, we provide the results obtained and accompanying discussion by 
employing the aforementioned graphlet orbits-based methodology, first with 
regards to the study of differentially expressed orbits (DEOs) and subsequently 
with regards to the classification of non-redundant OUPs by means of several 
classification models.

\subsection{Orbit enrichment and differential orbit expression analysis}
\begin{figure}[ht]
\centering
\includegraphics[scale=0.52]{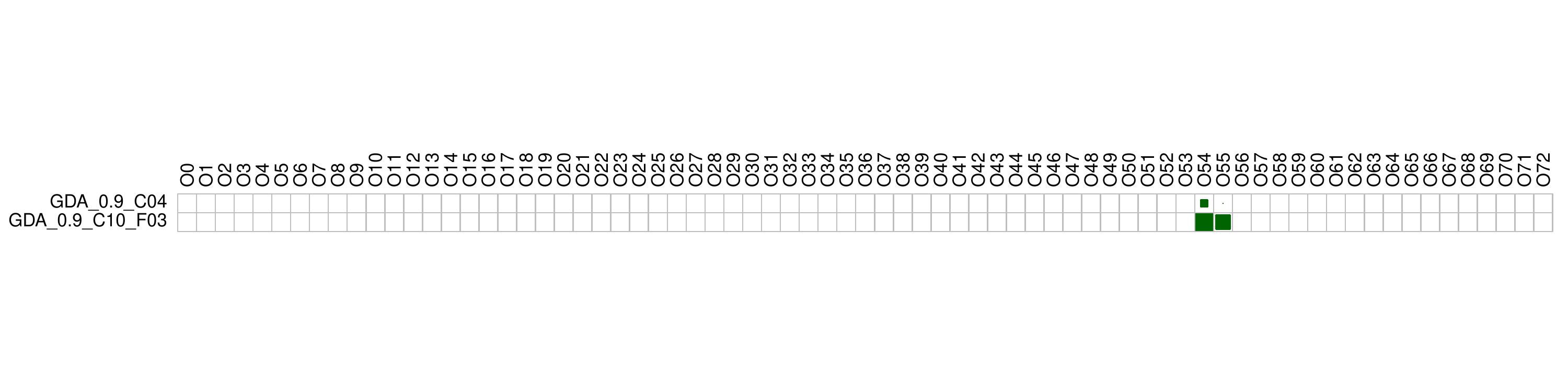}
\caption{Differentially expressed orbits (DEO): statistically significant orbits 
found to have \(Z-score > 2.58\) for at least \(50 \%\) networks of the same 
disease class.}
\label{fig:3_1_1}
\end{figure}

As mentioned before, the OUPs of disease networks were compared to the OUPs of 
two different kinds of random network ensembles. In the end, this leads to a 
matrix of networks with significantly overexpressed orbits. To illustrate, in 
the Figure \ref{fig:3_1_1}, we see the result of thresholding (Z-score) the 
differentially expressed orbit matrix (\emph{DEOM}) at \emph{Z-score} \(2.58\) 
or higher, i.e., 
keeping only the entries with \emph{Z-score} more than or equal to \(2.58\) and 
setting the others to zero. The corresponding output table is shown in Table S 
of the supplementary materials. Only orbit 54 is disproportionately represented 
in networks for neoplastic diseases, while orbits 54 and 55 are 
disproportionately represented in networks for NDDs. Differentially 
expressed orbits of these two classes are identical because they are equivalence 
groups of the same graphlet \(G_{22}\), which we refer to as a chevron.

\begin{table}[htbp]
\caption{Prediction performance of DNN on nrOUP features of two disease classes}
\centering
\begin{adjustbox}{max width=\textwidth}
\begin{tabular}{ r r r r r r r}
\hline
\multicolumn{1}{l}{Fold\#} & \multicolumn{1}{l}{AUPRC} & 
\multicolumn{1}{l}{Accuracy} & \multicolumn{1}{l}{Sensitivity} & 
\multicolumn{1}{l}{Speficity} & \multicolumn{1}{l}{Precision} & 
\multicolumn{1}{l}{MCC} \\ \hline
0 & 1.0000 & 0.9348 & 1.0000 & 0.8333 & 0.9032 & 0.8676 \\ 
1 & 0.9921 & 0.8696 & 0.8214 & 0.9444 & 0.9583 & 0.7483 \\ 
2 & 0.9893 & 0.9130 & 0.8929 & 0.9444 & 0.9615 & 0.8243 \\ 
3 & 0.9925 & 0.9565 & 1.0000 & 0.8889 & 0.9333 & 0.9108 \\ 
4 & 0.9879 & 0.9111 & 0.8571 & 1.0000 & 1.0000 & 0.8330 \\ 
\multicolumn{1}{l}{Average (Std)} & 0.9923 (0.0047) & 0.9170 (0.0323) & 0.9143 
(0.0822) & 0.9222 (0.0633) & 0.9513 (0.0359) & 0.8368 (0.0601) \\ \hline
\end{tabular}
\end{adjustbox}
\label{Tab:3_1_1}
\end{table}

\subsection{Performance of non-redundant OUP (nrOUPs) based DNN 
model}

\begin{figure}
\centering
\begin{subfigure}{.495\textwidth}
  \centering
  \includegraphics[width=\linewidth, right]{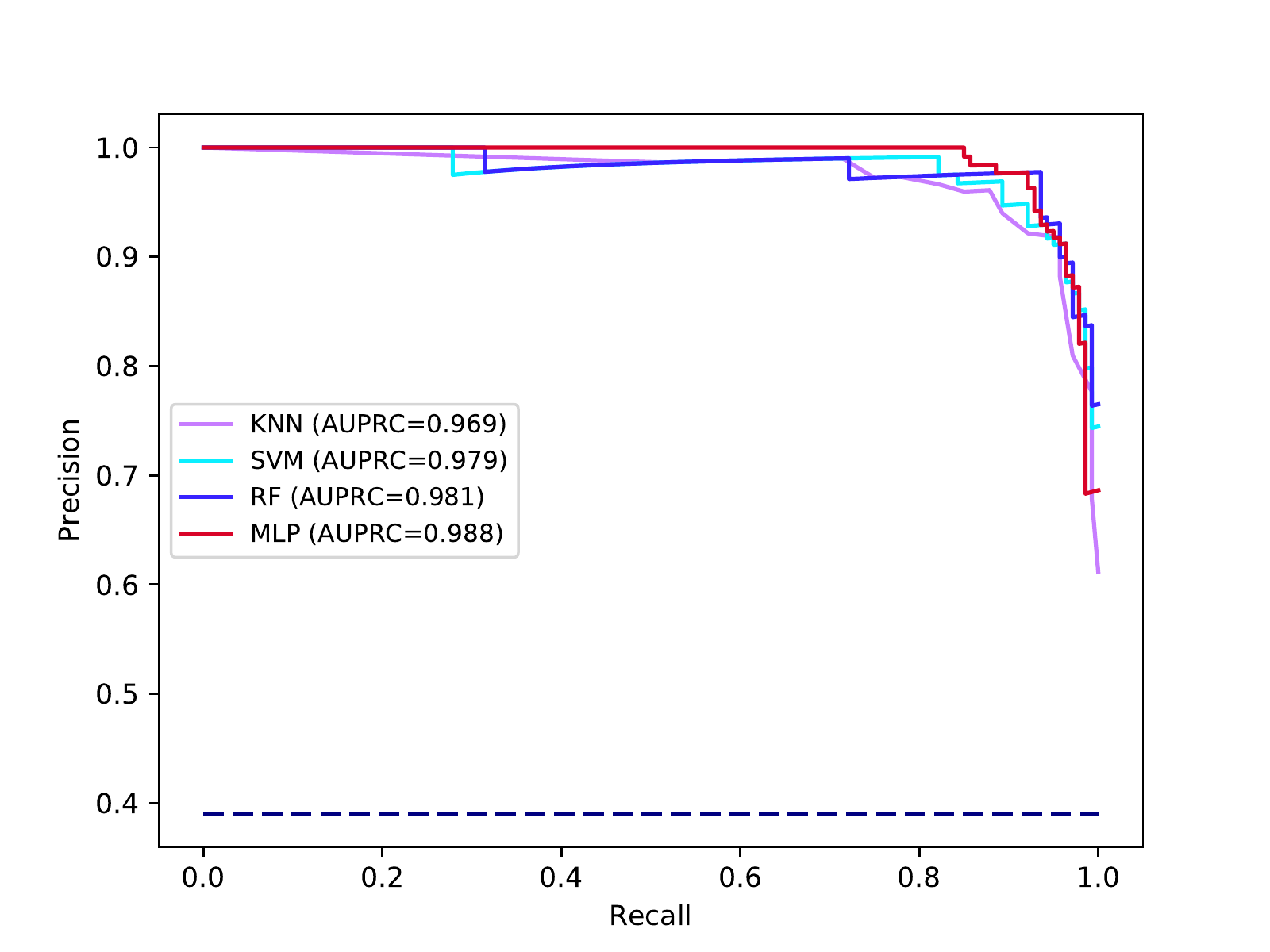}
  \caption{}
  \label{fig:3_1_2}
\end{subfigure} %
\begin{subfigure}{.495\textwidth}
  \centering
  \includegraphics[width=\linewidth, left]{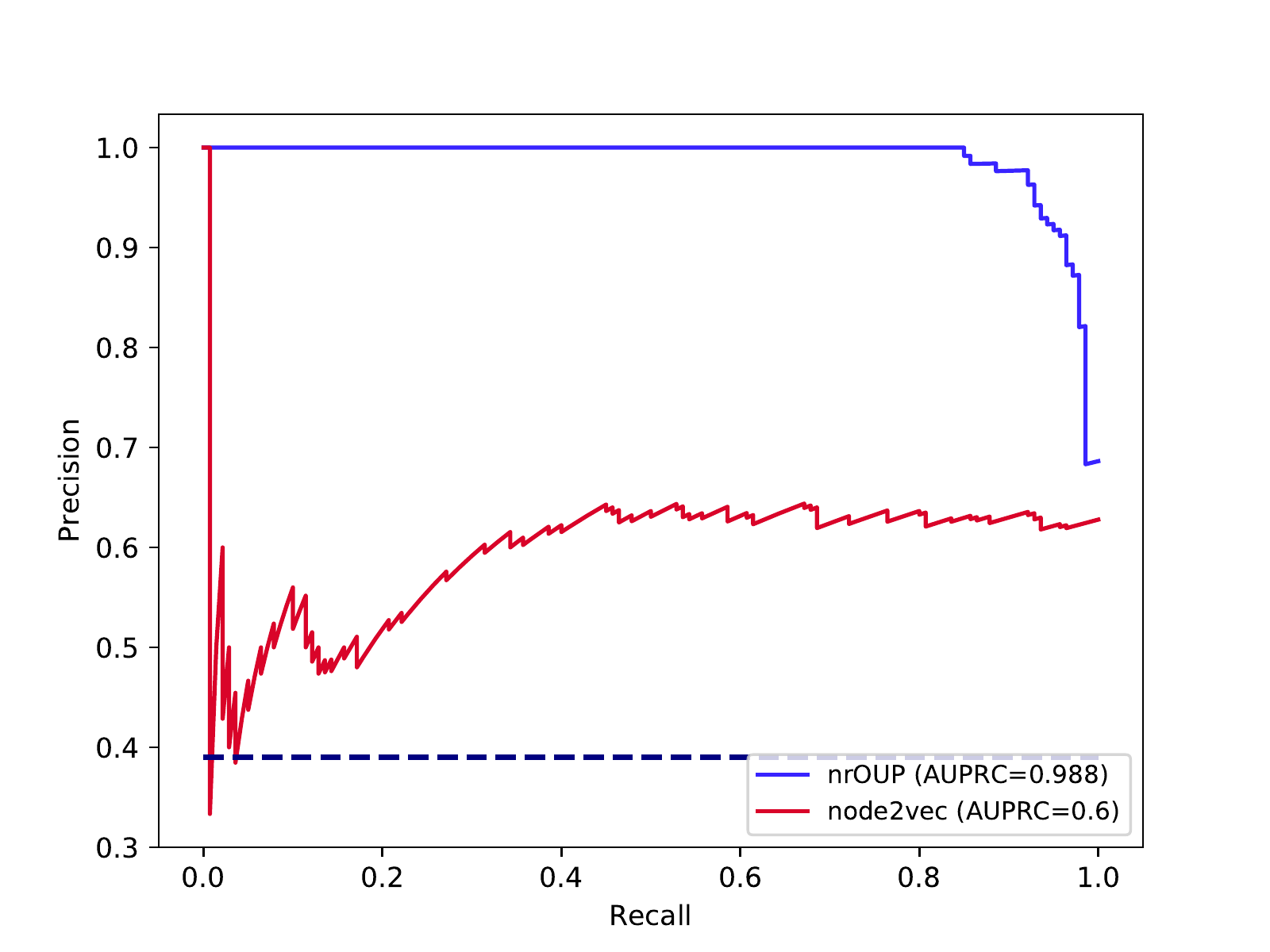}
  \caption{}
  \label{fig:3_1_3}
\end{subfigure}
\caption{Performance of various classifiers for their ability to predict disease 
networks (a). The deep neural network (DNN) model beat the K-nearest neighbour 
(KNN), support vector machine (SVM), and random forest (RF) classifiers on the 
independent test set, as measured by the areas under the PrecisionRecall curves 
(AUPRC). (b) How well nrOUP and node2vec, two separate node embedding 
techniques, perform in DNN classifiers. According to the AUPRC, the nrOUP 
technique is superior to the node2vec encoding scheme.}
\label{fig:auprc}
\end{figure}

Our earlier work shows that the DNN outperforms other ML algorithms on a local 
topology-based measure called the orbit usage profile for distinguishing across 
classes of networks. As a result, we used the count of orbits in graphlets to 
encode each network as a 72-dimensional OUP vector, from which we removed the 
redundant orbits to obtain 56 non-redundant OUPs (nrOUPs). nrOUPs were 
classified using a deep multi-layered neural network (DNN), while model 
parameters, such as the number and size of hidden layers, regularisation, 
activation function, and solver, were optimised using a five-fold 
cross-validation approach. In particular, the best results are achieved using 
four hidden layers of 1000 units each, the relu activation function, and the 
adam solver. In the training set, we saw an average prediction accuracy of 
\(97\%\), whereas, in the test set, it was only \(97\%\). The experimental 
findings on the nrOUPs datasets are summarised in Table \ref{Tab:3_1_1}, 
where we see that the 
average accuracy of the fivefold CV technique is \(91.7\%\), sensitivity is 
\(91.43 \%\), specificity is \(92.22 \%\), precision is \(95.13 \%\), the MCC 
is \(83.6 \%\), and the AUPRC value is \(0.9923\). Standard deviations for 
these performance measures are observed to be \(3.23, 8.22, 6.33, 3.59, 6.01\) 
and \(0.0047\), respectively. The lowest accuracy value recorded was 86.96
while the maximum accuracy value was up to \(95.65 \%\) across all five sets of 
predicted performance. We employed DNNs to categorise DEOs and compared the 
results to other popular classifier models. In particular, we optimised the 
hyperparameters of KNN, RF, and SVM models using five-fold cross-validation and 
trained them on the same nrOUPs features used to train DNN. Table 
\ref{Tab:3_1_2} summarises 
the experimental findings achieved by various classifiers on the nrOUPs dataset. 
Table \ref{Tab:3_1_2} shows that, compared to previous classifier models, the 
proposed model using the DNN as the classifier achieved much higher accuracy 
and AUPRC values (Figure \ref{fig:3_1_2}). Accuracy-wise, RF and DNN 
classifiers are on par; however, AUPRC for DNN is marginally superior to the RF 
classifier.

\begin{table}[htbp]
\caption{Prediction performance of different classifiers on nrOUP features of 
two disease classes}
\begin{adjustbox}{max width=\textwidth}
\begin{tabular}{l r r r r r r r r r r r r}
\hline
\multicolumn{ 1}{l}{\textbf{Classifier}} & \multicolumn{ 2}{c}{\textbf{AUPRC}} & 
\multicolumn{ 2}{c}{\textbf{Accuracy}} & \multicolumn{ 
2}{c}{\textbf{Sensitivity}} & \multicolumn{ 2}{c}{\textbf{Speficity}} & 
\multicolumn{ 2}{c}{\textbf{Precision}} & \multicolumn{ 2}{c}{\textbf{MCC}} \\ 
\cline{ 2- 13}
\multicolumn{ 1}{l}{} & \multicolumn{1}{c}{\textbf{Average}} & 
\multicolumn{1}{c}{\textbf{Std}} & \multicolumn{1}{c}{\textbf{Average}} & 
\multicolumn{1}{c}{\textbf{Std}} & \multicolumn{1}{c}{\textbf{Average}} & 
\multicolumn{1}{c}{\textbf{Std}} & \multicolumn{1}{c}{\textbf{Average}} & 
\multicolumn{1}{c}{\textbf{Std}} & \multicolumn{1}{c}{\textbf{Average}} & 
\multicolumn{1}{c}{\textbf{Std}} & \multicolumn{1}{c}{\textbf{Average}} & 
\multicolumn{1}{c}{\textbf{Std}} \\ \hline
KNN & 0.9793 & 0.0184 & 0.9039 & 0.0199 & 0.8786 & 0.0479 & 0.9444 & 0.068 & 
0.9639 & 0.0423 & 0.811 & 0.0392 \\ 
SVM & 0.9833 & 0.0144 & 0.9216 & 0.0499 & 0.9429 & 0.0407 & 0.8889 & 0.0878 & 
0.9312 & 0.0533 & 0.8361 & 0.1062 \\ 
DNN & 0.9914 & 0.0049 & 0.9344 & 0.0161 & 0.95 & 0.0598 & 0.9111 & 0.0745 & 
0.9467 & 0.0424 & 0.8684 & 0.0269 \\ 
RF & 0.9842 & 0.0138 & 0.9388 & 0.0099 & 0.9357 & 0.0466 & 0.9444 & 0.0556 & 
0.9655 & 0.0339 & 0.877 & 0.0195 \\ \hline
\end{tabular}
\end{adjustbox}
\label{Tab:3_1_2}
\end{table}

\subsection{Performance of nrOUP against node2vec embedding}

\begin{figure}
\centering
\begin{subfigure}{.495\textwidth}
  \centering
  \includegraphics[width=\linewidth]{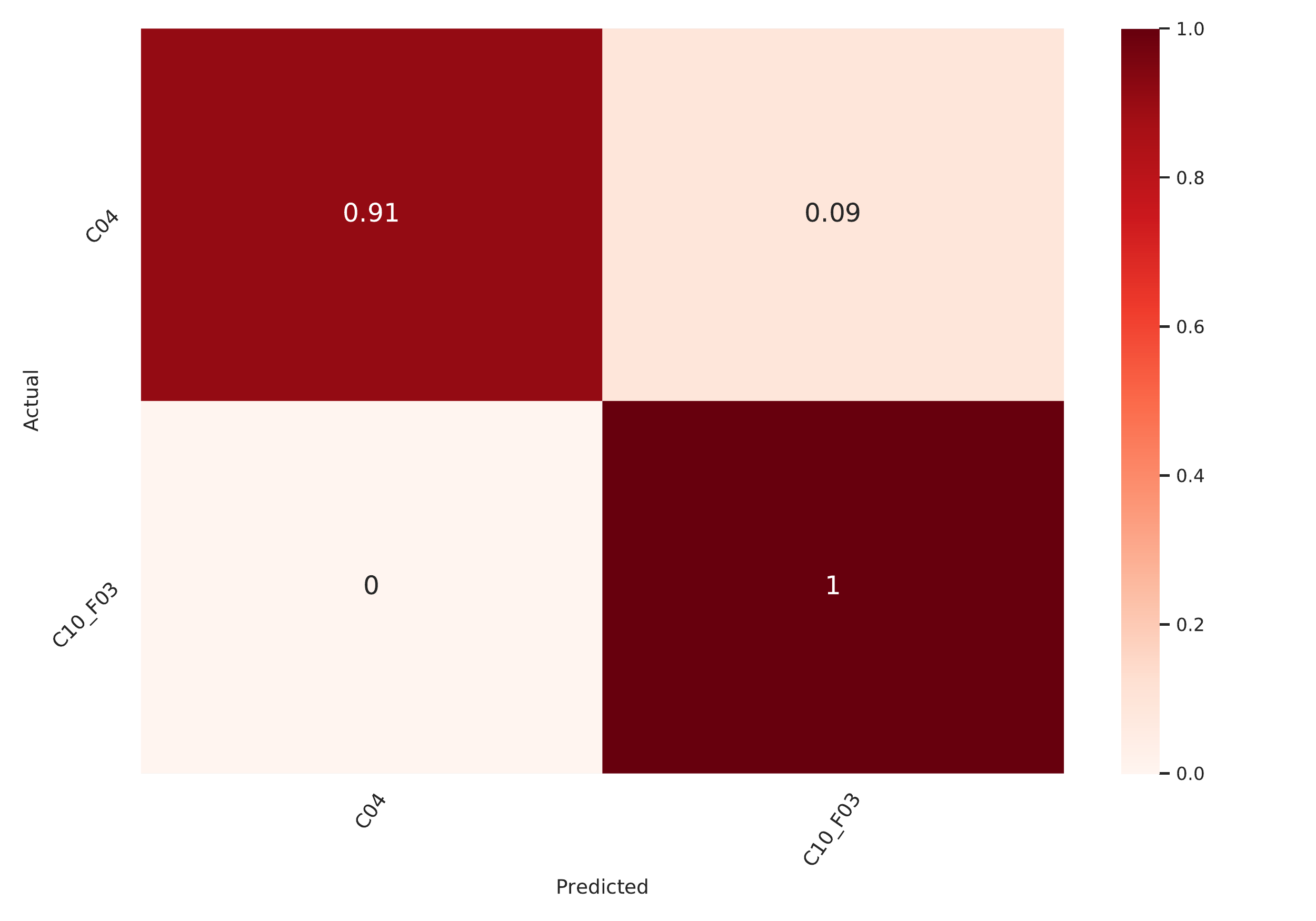}
  \caption{}
  \label{fig:3_1_4}
\end{subfigure}%
\begin{subfigure}{.495\textwidth}
  \centering
  \includegraphics[width=\linewidth]{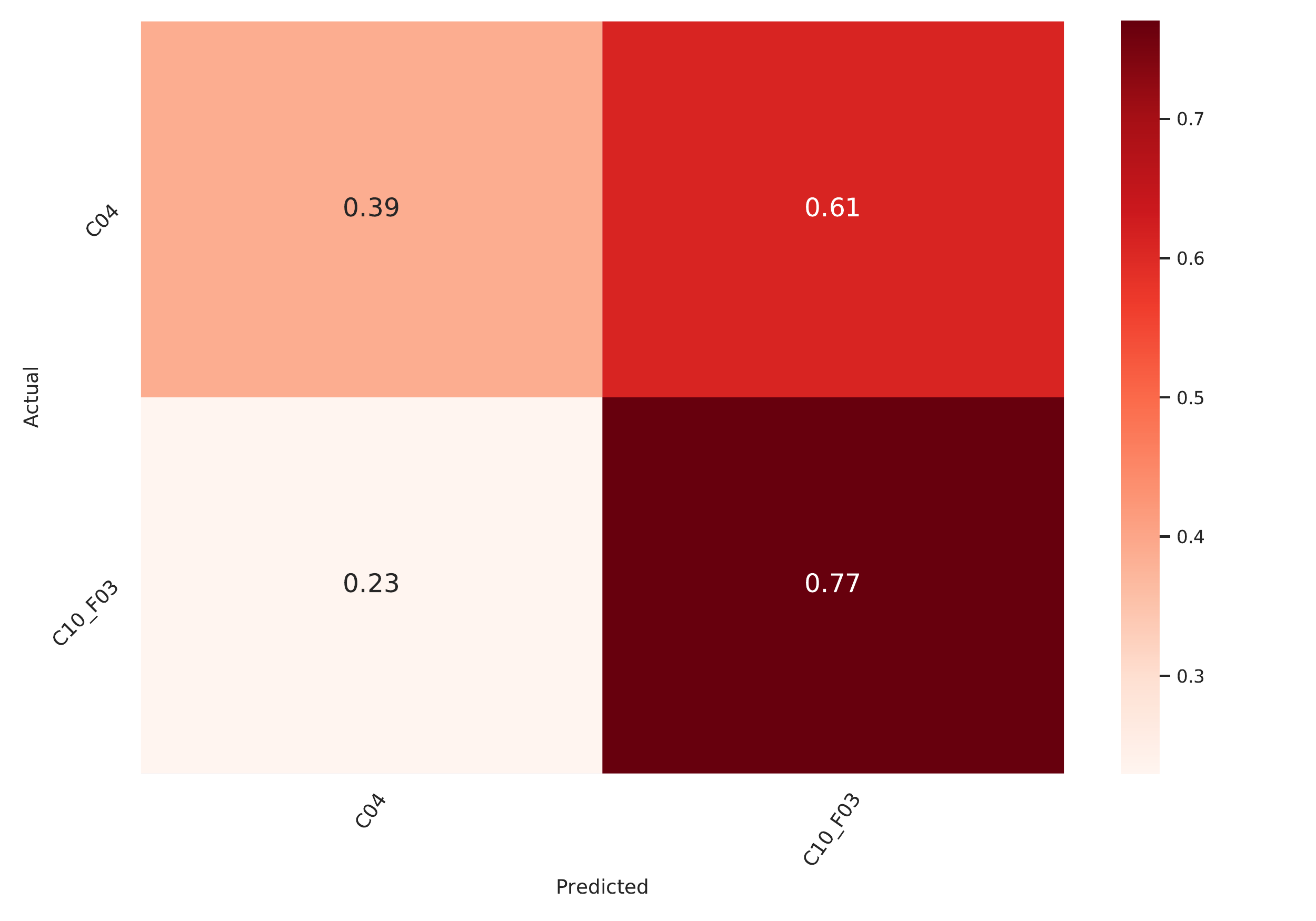}
  \caption{}
  \label{fig:3_1_5}
\end{subfigure}
\caption{Confusion matrices representing the performance of DNN on nrOUP (a) and 
node2vec (b) featurs.}
\label{fig:cm}
\end{figure}

Each network was transformed into a 66-dimensional vector representation, much 
like nrOUPs, by first generating a 12-dimensional vector representation via the 
node2vec technique. Here, we specify values 4, 10, and 12 for three 
hyperparameters, defining the walk length per source, the number of walks per 
source, and the output dimensions. We then employed cosine similarity to 
transform the data from each dimension into a (12, 12) matrix. Then, we took the 
lower triangular half of the similarity matrix (66 dimensions in total) and 
utilised it as a feature vector to characterise the whole network. Table 
\ref{Tab:3_1_3} displays the results of a classification exercise utilising DNN 
and node2vec embeddings. Compared to nrOUPs, the AUPRC obtained utilising 
node2vec features is extremely low, only 0.6 (Figure \ref{fig:3_1_3}). 
Moreover, nrOUPs improved upon all of the six performance evaluation metrics 
shown in Table \ref{Tab:3_1_3} compared to node2vec embeddings. We further 
generated confusion matrices for both nrOUP and node2vec features. Only nine per 
cent of the NP networks were misclassified into NDD networks when 
we 
used nrOUP features (Figure \ref{fig:3_1_4}). On the other hand, \(61 \%\) of 
neoplastic networks were predicted as neurological networks, and \(23 \%\) of 
neurological 
networks were misclassified as neoplastic networks (Figure \ref{fig:3_1_5}). 

\begin{table}[htbp]
\caption{Prediction performance of DNN on nrOUP and ndoe2vec features of two 
disease classes}
\begin{adjustbox}{max width=\textwidth}
\begin{tabular}{lcccccccccccc}
\hline
\multicolumn{1}{l}{\textbf{Features}} & \multicolumn{2}{c}{\textbf{AUPRC}} & 
\multicolumn{2}{c}{\textbf{Accuracy}} & 
\multicolumn{2}{c}{\textbf{Sensitivity}} 
& \multicolumn{2}{c}{\textbf{Speficity}} & 
\multicolumn{2}{c}{\textbf{Precision}} & \multicolumn{2}{c}{\textbf{MCC}} \\ 
\cline{ 2- 13}
\multicolumn{ 1}{l}{} & \textbf{Average} & \textbf{Std} & \textbf{Average} & 
\textbf{Std} & \textbf{Average} & \textbf{Std} & \textbf{Average} & 
\textbf{Std} 
& \textbf{Average} & \textbf{Std} & \textbf{Average} & \textbf{Std} \\ \hline
nrOUP & 0.9919 & 0.0048 & 0.9257 & 0.0251 & 0.9286 & 0.0714 & 0.9222 & 0.0843 & 
0.9533 & 0.0485 & 0.8535 & 0.0478 \\
Node2vec & 0.6499 & 0.1239 & 0.5722 & 0.0576 & 0.6786 & 0.0253 & 0.4052 & 
0.1706 
& 0.6484 & 0.0677 & 0.0814 & 0.154 \\ \hline
\end{tabular}
\end{adjustbox}
\label{Tab:3_1_3}
\end{table}

\section{Summary and Conclusions}

Cancer and other brain-related disorders account for a disproportionate share of 
all disease-related deaths. In a recent study called the Pan-Cancer Initiative, 
researchers found that while tumours in the same tissue can have widely 
different genetic traits, tumours in other organs share common molecular 
features. Those findings prompted us to question whether or not we could use 
network-level similarities and differences to categorise diseases into their 
respective classes. As part of our research, we built 287 networks for two types 
of diseases: tumours (112) and neurological disorders (175). The networks' local 
architectures were then investigated to collect individual profiles of orbital 
usage called OUPs. Orbit 54 was shown to be differentially expressed in 
neoplastic networks, while orbits 54 and 55 were found to be differentially 
expressed in brain illness networks. Following this, 56 unique OUPs were 
extracted from the OUPs and used as features for disease network classification. 
After training four classifiers (KNN, SVM, RF, and DNN) with nrOUPs, DNN proved 
to be the most effective and had the highest average AUPRC of 0.988. Finally, we 
evaluated the performance of DNN on nrOUPs and node2vec embeddings and found 
that nrOUP performed better than node2vec in all the six performance evaluation 
metrics used in this study.

\bibliographystyle{Science}
\bibliography{GDA}

\section*{ACKNOWLEDGEMENTS}
$\text{VS}^{\dagger}$ thanks Council of Scientific and Industrial Research 
(CSIR), India for providing Junior Research Fellowship (JRF). \textbf{Funding:} 
Authors recieved no specific funding for this research work. \textbf{Authors 
Contributions:} $\text{VS}^*$ conceptualized and designed the research 
framework. $\text{VS}^{\dagger}$ performed the computational 
experiments. $\text{VS}^{\dagger}$ and $\text{VS}^*$ analyzed the data and 
interpreted results. $\text{VS}^{\dagger}$ and $\text{VS}^*$ wrote and 
finalized the manuscript. \textbf{Competing Interests:} The authors declare 
that they have no conflict of interests. \textbf{Data and materials 
availability:}

\end{document}